# EXOSAT observations of the ultra–soft X–ray binary 4U1957+11

D. Ricci[1,4], G.L. Israel[2,4], and L. Stella[3,4]

[1] Dipartimento di Fisica, Università di Roma "La Sapienza", P.le Aldo Moro 5, I–00185 Roma, Italy, dona@vxrmg9.icra.it
[2] International School for Advanced Studies (SISSA–ISAS), Via Beirut 2-4, I–34014 Trieste, Italy, israel@vxrmg9.icra.it
[3] Osservatorio Astronomico di Brera, Via E. Bianchi 46, I–22055 Merate, Italy, stella@astmim.mi.astro.it
[4] Affiliated to I.C.R.A.

**Abstract.** We present results from the analysis of the two *EXOSAT* observations of the low mass X-ray binary (LMXRB) 4U1957+11. The 1-20 keV spectrum of the source is best fitted by a power law model with exponential cutoff, that provides an approximation of a thermal Comptonisation spectrum. The cutoff energy ($\sim 2$ keV), as well as the X-ray colours, are intermediate between those of ultrasoft sources containing black hole candidates (BHCs) and those of soft LMXRBs containing an old accreting neutron star. We find no evidence for a black body spectral component. During the 1985 observation the source flux was $\sim 10\%$ higher than in the 1983 observation, and the ME spectra provide evidence for a high energy tail (up to $\sim 16$ keV), which strengthens the similarity with the spectra of high state BHCs. However the source luminosity was only $\sim 5 \times 10^{36}$ ergs s$^{-1}$.

During the 1985 observation the GSPC spectra revealed the presence of a 100 eV EW iron K-line emission feature with a centroid energy of $7.06 \pm 0.09$ keV; this value is among the highest measured by *EXOSAT*. The ME light curves showed only a low-amplitude variability on timescales of a few hours. No evidence for a periodic modulation with periods between 0.016 and 13000 s was found.

## 1. Introduction

4U1957+11 is a relatively poorly studied $20 - 80\mu$Jy X-ray source close to the galactic plane. Discovered with *Uhuru* (Giacconi et al. 1974) the source was subsequently associated with a blue star-like optical counterpart (V=18.7, B-V=0.3 and U-B=-0.6), thanks to the SAS-3 positioning capabilities (Margon et al. 1978). This is located within 10 arcsec of the position of 4U1957+11 and has colours similar to those of the optical counterpart of Sco X-1 and 4U0614+09. Optical CCD photometry of 4U1957+11 revealed a $\sim$0.2 mag roughly sinusoidal modulation with a period of 9.3 hr, which most likely corresponds to the orbital period of the system (Thorstensen

*Send offprint requests to*: L.Stella

1987). These findings clearly indicate that 4U1957+11 is a low mass X-ray binary (LMXRB) with an X-ray luminosity of a few $\times 10^{36}$ ergs s$^{-1}$ for the estimated distance of $\sim$6 kpc.

X-ray data were also obtained with the *Vela 5B* and *HEAO-1* satellites. No significant long-term periodicity between 20 d and 1 yr was found in the *Vela 5B* data (Priedhorsky & Terrel 1984). In the *HEAO-1* A-2 X-ray colour-colour diagram assembled by White and Marshall (1984) 4U1957+11 lies close to the region occupied by ultrasoft sources, most of which are BHCs in their high state, such as Cyg X-1, LMC X-3, A0620-00 and GX 339-4. A similar conclusion is obtained based on the *EXOSAT* X-ray colour-colour diagram (Schultz et al. 1989). These results suggest that 4U1957+11 might host a BHC. However, in a recent paper Singh et al. (1994) report that the *EXOSAT* spectra of 4U1957+11 are similar to those of high luminosity LMXRBs containing an accreting weakly magnetic neutron star. Based on a $\sim$ 33 hr Ginga observation, Yaqoob et al. (1993) conclude that the source spectrum is well represented by the sum of a soft component, which they model with a black-body emitting disk, and a hard component (with a power law photon index of $\sim$ 2) detected up to energies of $\sim$ 18 keV. The derived values of the inner radius and temperature of the disk are $r_{in}(\cos i)^{1/2} \simeq 2$ km ($i$ is the inclination angle) and $kT_{in} \simeq 1.5$ keV, respectively, suggesting a similarity with LMXRBs containing a neutron star.

In this paper we present a detailed analysis of the *EXOSAT* spectral and timing data. Details of the *EXOSAT* observations are summarised in Section 2. Section 3 and 4 describe our analysis of the source spectra and light curves, respectively. Our results are discussed in Section 5.

## 2. The *EXOSAT* observations of 4U 1957+11

The European Space Agency's X-ray observatory *EXOSAT* was operational from May 1983 to April 1986. Its *low energy imaging telescopes* (LE1 and LE2), used in conjunction with the *channel multiplier array* (CMA), covered the 0.05-2 keV band and provided broad band filter spectroscopy. The Argon chambers of the *medium energy* (ME) proportional counter array operated over the 1-20 keV band and produced spectra (with a resolution of $\Delta E/E \sim 0.2 \, (E/7 \text{ keV})^{-1/2}$) and light curves with high throughput. The ME was often operated with one half of the detector array offset from the source in order to provide a simultaneous monitor of the particle background; the pointing directions of the two halves were usually swapped every 3–4 hours in order to minimize the systematic uncertainties in the background subtraction. The *gas scintillation proportional counter* (GSPC) provided a factor of $\sim 2$ improved spectral resolution in the 2-20 keV band, with a factor of $\sim 5$ lower effective area than the ME.

**Table 1.** *EXOSAT* observations and count rates

| Time | Instrument | Filter or Energy Range | Exposure (s) | Counts s$^{-1}$ |
|---|---|---|---|---|
| 1983 | CMA-LE2 | 3000 Å Lexan | 6203 | 0.48±0.02 |
| August 27 | CMA-LE2 | Aluminium/Parylene | 5814 | 0.34±0.01 |
| | CMA-LE2 | Boron | 6430 | 0.21±0.01 |
| | CMA-LE2 | Polypropylene | 536 | 0.46 ±0.04 |
| | ME | 1–20 keV | 20282 | 66.9±0.2 |
| | GSPC | 3–10 keV | 26654 | 4.69±0.02 |
| 1985 | CMA-LE1 | 3000 Å Lexan | 15228 | 0.42±0.01 |
| May 28–29 | CMA-LE1 | Alumimium/Parylene | 10201 | 0.31±0.01 |
| | CMA-LE1 | Boron | 11954 | 0.18±0.007 |
| | ME | 1–20 keV | 37265 | 80.18±0.19 |
| | GSPC | 3–10 keV | 36897 | 5.19±0.02 |

*EXOSAT* observed 4U1957+11 on 1983 August 27 for about 6 hr and on 1985 May 28–29 for about 11 hr. The data were obtained from the *EXOSAT* database available within High Energy Astrophysics Database Service at the Astronomical Observatory of Brera (Tagliaferri & Stella 1993) except for the high time resolution light curves presented in Section 4, which were accumulated by using ESA's on-line interactive analysis system. Table 1 gives a summary of the source count rate and exposure times in each instrument during the two observations. In all cases the instrumental background was stable.

## 3. Energy Spectra

### 3.1. ME and LE data

The 1–20 keV pulse height source spectra from the ME Argon chambers were analysed together with the source count rates derived from the LE-CMA and each of the filters used. The data from the ME Xenon chambers were excluded due to systematic uncertainties in the background subtraction. Several trial spectral models were convolved through the instrumental response and fitted to the data. Neither a single power law, a thermal bremsstrahlung, nor a black body model produced an acceptable representation of the spectrum of 4U1957+11.

A power-law model with an exponential high-energy cutoff ("cutoffpl") provided instead a good fit to the 1983 data, for a power-law photon index of $\Gamma_S \sim 0.2$ and a cutoff energy of $E_{cut} \sim 2.1$ keV (Table 2 and Fig. 1). The column density was constrained to a value of $(5.2 \pm 0.8) \times 10^{20}$ H cm$^{-2}$ mainly through the LE rates measurements.

The thermal Comptonisation model of Sunyaev & Titarchuck (1980) ("CompST"), of which the power law with exponential cutoff model provides an approximation, gave instead a substantially worse fit. In this case an electron temperature of $kT_e \sim 1.4$ keV and a Thomson optical depth of $\tau_{es} \sim 22.1$ were estimated. For both models the 1-20 keV luminosity of the source was derived to be $L_X \simeq 4.6 \times 10^{36}$ ergs s$^{-1}$.

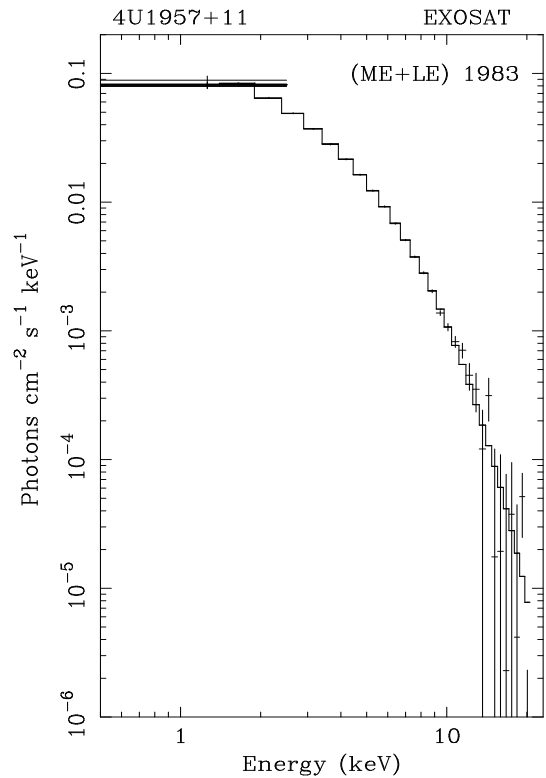

**Fig. 1.** ME and LE spectrum from the 1983 observation; the "cutoffpl" model has been used.

We note that the integrated spectrum of an accretion disk locally emitting like a black body provided also a good fit to the 1983 data. However, the standard $\alpha$-viscosity disk model ("disk" in table 2 and 3; cf. Shakura & Sunyaev 1973) requires a disk luminosity of $5.1 \times 10^{39}$ ergs s$^{-1}$ or $\sim 10$ times the Eddington limit for the estimated mass $\sim 4$ M$_\odot$ of the accreting black hole. Reconciling this luminosity with the observed $L_X$ requires an unrealistic inclination of about 89.9° and the presence of X-ray eclipses, which are excluded, e.g., by the 1985 EXOSAT light curves. A similarly good fit was obtained by using the simple black-body disk model adopted by Yaqoob et al. 1993 ("diskbb" in Table 2 and 3; see also Mitsuda 1984), in which the disk inner radius and temperature are allowed to freely vary. The best fit values, $r_{in}(\cos i)^{1/2} \simeq 1.9$ km and $kT_{in} \simeq 1.6$ keV, are similar to those obtained with the Ginga spectra. In turn a very high inclination ($i > 87.7°$) would be required if the derived inner disk radius were to be compatible with the neutron star radius or the marginally stable orbit of a few solar masses black hole.

**Table 2.** Spectral fits of the 1983 data. Errors in the spectral parameters are at 68% confidence level.

| Date | Model | $\chi^2/dof$ | Best fit parameters |
|---|---|---|---|
| 1983 | **cutoffpl** | 21.6/28 | $\Gamma_S = 0.18 \pm 0.04$, $E_{cut} = 2.07 \pm 0.04\ keV$ |
| Aug. 27 | **compST** | 65.2/28 | $kT = 1.37 \pm 0.01\ keV$, $\tau_{es} = 22.1 \pm 0.4$ |
| | **disk** | 30.9/28 | $M = 3.83\ M_\odot$, $\dot{M} = 9.9\ \dot{M}_{Edd}$, norm$^a$= $1.3 \times 10^{-3}$ |
| | **diskbb** | 36.2/29 | $T_{in} = 1.63 \pm 0.01\ keV$, , norm$^b$= 8.02 |

$^a$ norm=$\cos(i)/d_{10}^2$, where $i$ is the inclination of the disk and $d_{10}$ is the distance in units of 10 kpc.
$^b$ norm=$(R_{in}/d_{10})^2 \cdot \cos(i)$, where $R_{in}$ in Km, is the inner disk radius.

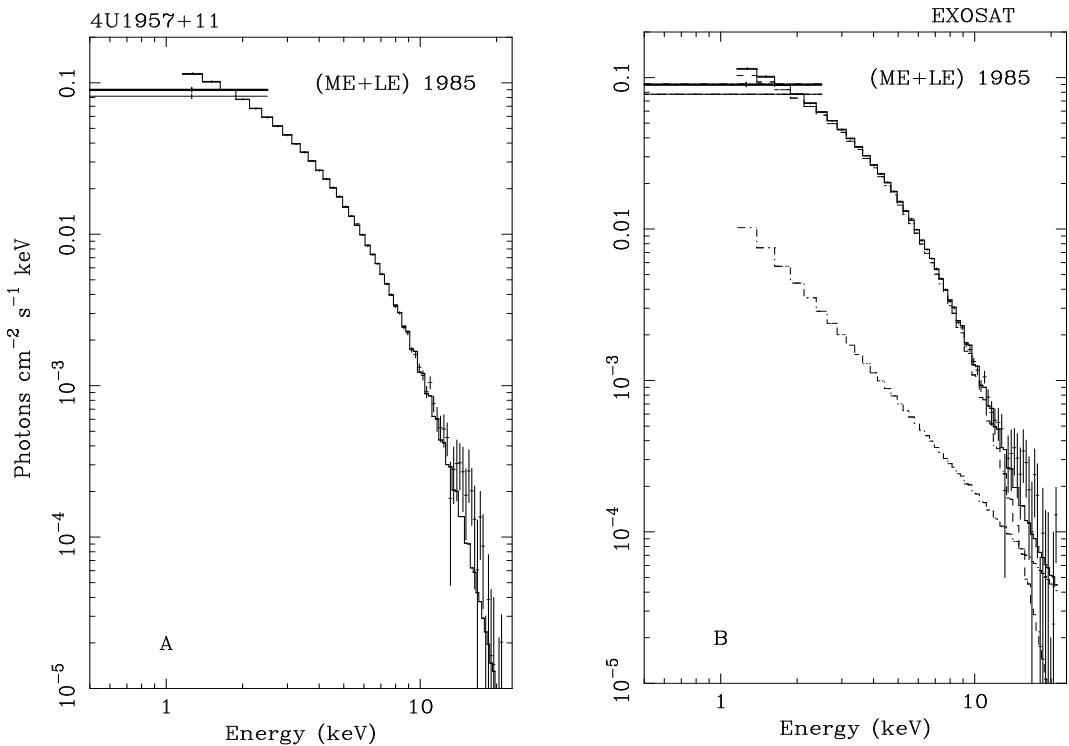

**Fig. 2.** ME and LE spectrum from the 1985 observation. In panel A the single component "cutoffpl" model has been used. In panel B a second power-law component with a slope of 2 has been added in order to fit the high energy excess.

The source count rate increased by $\sim 20\%$ during the 1985 observation. The values obtained from the spectral fitting of the 1985 data are similar to those discussed above (Table 2), except for a rather noticeable high energy-excess in the ME rates above 10 keV. This excess was not adequately modelled by any of the single component models adopted for the 1983 observation. The cutoff power law provided the best single component model with a $\chi^2$ of 54.8 for 57 degrees of freedom ($dof$) (see Fig. 2 A). To model the high-energy excess an additional spectral component consisting of a power law was added. Owing to poor statistics, the photon index of this high energy power law could not be obtained with an acceptable accuracy (range between −1.2 and 2.5). Therefore we fixed its value to 2, compatible with the measurement by Ginga (Yaqoob et al. 1993). A $\chi^2/dof$ of 48.9/56 was obtained in this case. The statistical significance of the additional power law component was evaluated through an F-test to be $\sim 10^{-2}$ (Fig. 2 B). Therefore we conclude that the 1985 EXOSAT data confirm the Ginga result of an additional spectral component above $\sim 10$ keV. The 1-20 keV source luminosity during the 1985 observation was $L_X \simeq 5.1 \times 10^{36}$ergs s$^{-1}$, with the high energy component contributing up to $\sim 10\%$ of the total.

### 3.2. GSPC observations

The 3–10 keV GSPC spectra from both observations were fitted with the same continuum model that provided the best fit to the ME and LE data (note that no evidence for a high energy excess is found in the ME data below 10 keV). The photoelectric absorption was kept constant at the best fit value obtained from the ME and LE data. The best fit model gave a $\chi^2/dof$ of 120/100 and 138.8/97 for the 1983 and 1985 observations, respectively (Fig. 3).

**Table 3.** Spectral fits of the 1985 data.

| date | model | $\chi^2/dof$ | Best fit parameters |
|---|---|---|---|
| 1985 | **cutoffpl** | 54.8/57 | $\Gamma_S = 0.19 \pm 0.03$, $E_{cut} = 2.08 \pm 0.03\ keV$ |
| May 28–29 | **compST** | 137/57 | $kT = 1.37 \pm 0.01\ keV$, $\tau = 22.13 \pm 0.26$ |
| | **disk** | 72.8/56 | $M = 3.8\ M_\odot$, $\dot{M}=9.9\ \dot{M}_{edd}$, norm$= 1.5 \times 10^{-3}$ |
| | **diskbb** | 83.5/58 | $T_{in} = 1.63 \pm 0.01\ keV$, , norm$= 9.01$ |

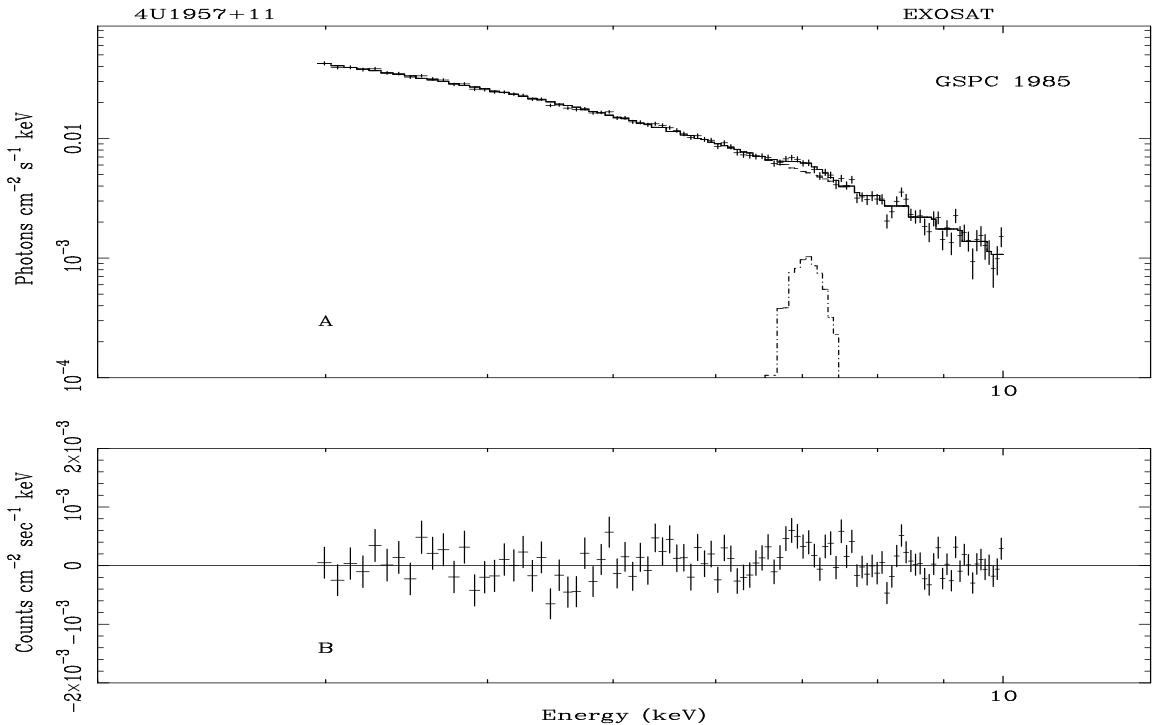

**Fig. 3.** GSPC spectrum from the 1985 observation. Panel A shows the fit using the "cutoffpl" model plus a Gaussian line. Panel B shows the residuals with respect to the continuum.

Adding a Gaussian line profile with a centroid energy in the 6–7 keV range did not improve significantly the fit of the 1983 GSPC spectrum. The 90% confidence upper limit to the equivalent width of such a feature was $\sim 90$ eV. On the contrary, adding a Gaussian line profile decreased the $\chi^2/dof$ of the 1985 GSPC spectra to 120.1/95, corresponding to an F-test probability of $\sim 0.001$ (see also Gottwald & White 1990). The centroid energy and equivalent width of the line were determined to be $7.06 \pm 0.09$ keV and $102 \pm 23$ eV, respectively. The width of the line could not be resolved by the GSPC and an upper limit of 0.9 keV (90% confidence) was obtained for the FWHM. We note that by adopting the alternative X-ray continuum models discussed in Sect. 3.1, the derived line parameters remain essentially unchanged.

## 4. Time variability

### 4.1. Light curves

During the observations of 1983 and 1985 the ME provided respectively 16 and 64 channel pulse height analyser data from both the source and background detectors. Using these data, lightcurves with 0.3125, 0.625 and 30 s time resolution in the energy ranges 1–4 keV, 4–10 keV and 1–10 keV were extracted from the *EXOSAT* database. The lightcurves display a small amplitude aperiodic variability on long timescales which is more pronounced in the 1985 observation (see Fig. 4).

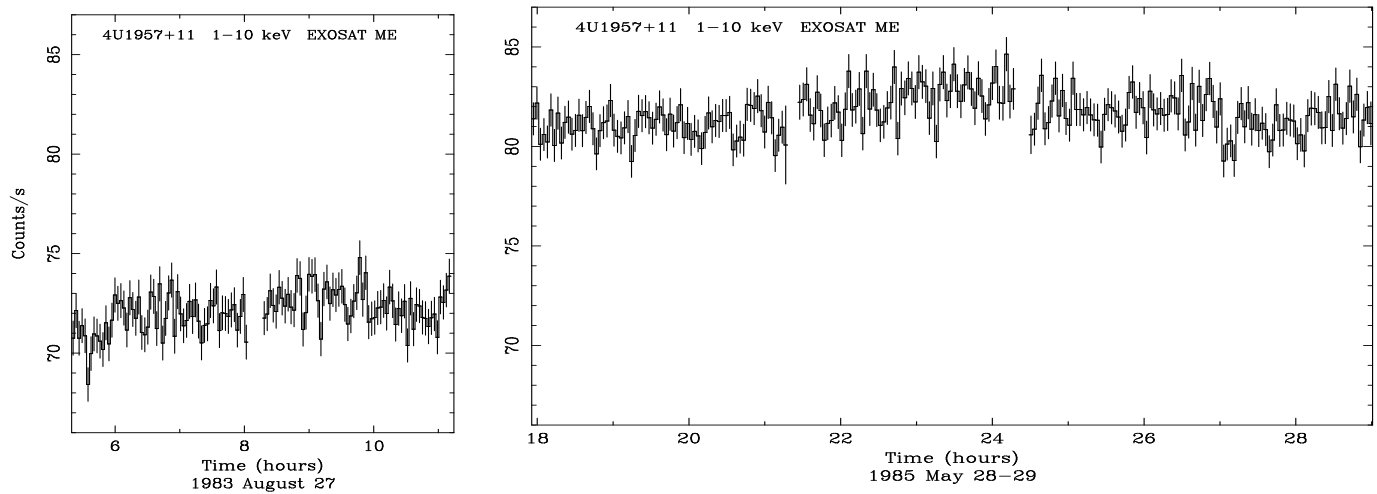

**Fig. 4.** 1–10 keV ME lightcurves from both observations. Each bin corresponds to an integration time of 166 s

## 4.2. Power Spectra

The lightcurves described above were used to carry out a detailed search for coherent pulsations. Since the 1985 observation is longer and has a higher average count rate than the 1983, it allows for a more sensitive search. In order to maximize the Fourier resolution, and therefore the sensitivity, a single power spectrum was calculated for the 0.3125, 0.625 and 30 s resolved lightcurves of each observation and energy range. Missing data points were replaced with the average count rate. In Fig. 5 the power spectra are marked with "a" and normalised such that the white noise arising from counting statistics corresponds to a power of 2.

A red noise component is dominant in the frequency range from $\sim 5 \times 10^{-5}$ Hz to $\sim 3 \times 10^{-4}$ Hz. This component results from the long–timescale variability visible in the lightcurves. The presence of this continuum power spectrum component affects the statistical distribution of the power estimates for long periods. This means that techniques which assign probabilities to power spectrum peaks assuming the simple $\chi^2$ distribution of power expected from counting statistics noise (see e.g. Leahy et al. 1983) may significantly overestimate their significance. Israel & Stella (1995) developed a technique for reliably searching for power spectrum peaks, corresponding to a periodic modulation, in the presence of "coloured" continuum power spectrum components. A search for coherent pulsations is carried out by looking for significant peaks above the estimated continuum spectrum. If no peaks with a probability of chance occurrence lower than a given threshold are found, then an upper limit to the semi-amplitude of a sinusoidal modulation is worked out for each Fourier frequency of the power spectrum through a generalisation of the method outlined by Leahy et al. (1983). By using this technique, each of the power spectra of 4U1957+11 was searched for peaks with a probability of $\leq 0.3\%$ ($3\sigma$ confidence level) of exceeding by chance the level of the continuum power spectrum components

For the 11 hr 1985 May 28–29 observation, power spectra were calculated from the 0.625 and 30 s time resolved lightcurves in the 1–10, 1–4 and 4–10 keV bands. No significant peaks were found in any of the power spectra. The 99.7% confidence upper limits to the modulation semi-amplitude are given in Table 4 for selected periods.

**Table 4.** $3\sigma$ upper limits to the semi-amplitude of pulsations for selected periods

| Period (s) | Upper limits 1983 August 27 | | | 1985 May 28–29 | | |
|---|---|---|---|---|---|---|
| | 1–10 | 1–4 | 4–10 | 1–10 | 1–4 | 4–10 |
| 13650 | – | – | – | 11.8 | 6.69 | 16.8 |
| 10240 | – | – | – | 7.71 | 4.13 | 11.2 |
| 8192 | 9.47 | 15.3 | 8.41 | 2.91 | 8.41 | 8.02 |
| 4096 | 4.52 | 6.32 | 3.27 | 0.83 | 0.85 | 1.52 |
| 512 | 0.55 | 0.69 | 0.81 | 0.41 | 0.52 | 0.59 |
| 16[a] | – | – | – | 0.67 | – | – |
| 8–0.03[a] | – | – | – | $\leq 0.85$ | – | – |
| 0.02[a] | – | – | – | 0.99 | – | – |
| 0.016[a] | – | – | – | 1.17 | – | – |

[a] From the high time resolution ME data (1–20 keV).

For short periods, up to $\sim 4000$ s, the sensitivity of the search is mainly limited by counting statistics noise and upper limits of $\leq 1\%$ are obtained for any sinusoidal modulation. For longer periods, the source red noise component plays an increasingly important role in reducing the sensitivity of the search. The corresponding upper limits have values of $\sim$ 3–17% for periods between 8000 and 14000 s. The *EXOSAT* observations were not long enough to search for a low amplitude X–ray modulation at the 9.3 hr orbital period.

We analysed also the power spectra of the 0.3125 and 30 s resolved 1–10, 1–4 and 4–10 keV ME light curves from the $\sim 6$ hr 1983 August 27 observation. All the points in the power spectra are well below the detection threshold and thus no significant periodicities are found. Upper limits for selected frequencies are given in Table 4 and shown in Fig. 6 (for the 1–10 keV energy range).

ME count rates from the entire energy range of the Ar chambers (1–20 keV) were also available with a resolution of

**Fig. 5.** Power spectra (a) of the ME lightcurve obtained from the 1–10 keV data (panel A) and from the high resolution data in the entire range of the ME Ar chambers (panel B) during the 1985 May 28–29 observation. The 99.7% confidence threshold for the detection of a sinusoidal signal is shown (b). The bottom curves (c) give the corresponding upper limits to the fractional semi-amplitude of the modulation.

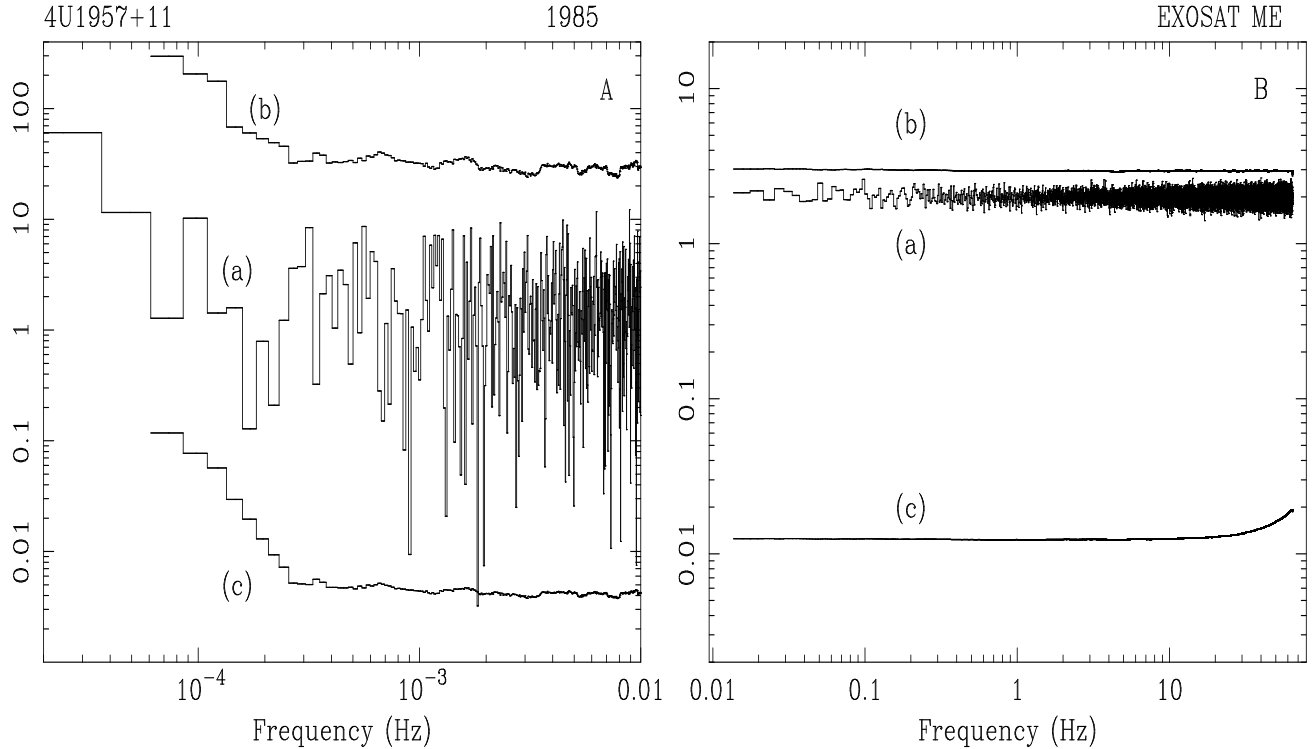

∼ 8 ms for the 1985 May 28–29 observation. We divided the lightcurve in 301 intervals, in order to prevent the Doppler frequency variations associated to the orbital motion from smearing the power of a sinusoidal signal over more than one Fourier frequency. These intervals were used to calculate 301 power spectra of 8192 indipendent Fourier frequencies.

The power spectrum resulting from the average of these power spectra was then analysed with the same technique discussed above. No significant periodicities were detected. The upper limits are reported in Table 4 and plotted in Fig. 5.

## 5. Discussion

The main component of the *EXOSAT* spectra is consistent with a power law model with exponential cutoff characterised by a spectral index of ∼ 0.2 and a cutoff energy of ∼ 2 keV for both the 1983 and 1985 observations. This spectral form provides a simple analytical approximation to the thermal Comptonisation spectrum of Sunyaev & Titarchuk (1980) and, besides semplicity, it involves no conceptual difference. Indeed thermal Comptonisation models (and their approximations) are found to provide very good fits to the main spectral component of a number of high state BHCs and LMXRBs containing old weakly magnetic neutron stars (White et al. 1988).

In high state BHCs this spectral component is somewhat softer then that of 4U1957+11 ($E_{\rm cut} \sim 1.4$ keV in LMC X–3 and LMC X–1; Ricci 1995). In LMXRBs containing an old neutron star the component is instead harder (e.g. $E_{\rm cut} \sim 7$ keV in 4U1705-44 and $E_{\rm cut} \sim 5$ keV in 4U1636-53; Ricci 1995). A classification of the compact object in 4U1957+11 based on the characteristic of its Comptonised spectral component is therefore premature.

On the other hand the 1985 *EXOSAT* spectra and, especially, the Ginga spectra show the presence of a high energy power-law component above energies of 10 keV. This kind of two-component spectra has been observed from several BHCs in their high state, the luminosity of which, however, is a factor of > 10 higher than that of 4U1957+11.

Our decomposition of the 1983 *EXOSAT* spectrum of 4U1957+11 is similar to that of Singh et al. (1994), who in their analysis used also the ME Xenon chambers data and excluded the LE-CMA rates. However, according to these authors, the 1985 *EXOSAT* spectrum of 4U1957+11 consists of the sum of a ∼ 1 keV temperature black body, accounting for about 40% of the total luminosity, plus a Comptonised spectrum extending up to ∼ 20 keV. In this case 4U1957+11 would posses an X-ray spectrum similar to that of the high luminosity LMXRBs (> $10^{37}$ ergs s$^{-1}$) containing an old neutron star (the so-called "Z–sources")(White et al. 1988, Hasinger & van der Klis 1989). Such a spectral decomposition of the 1985 *EXOSAT* data appears however unlikely, for the following reasons: (a) despite the moderate (∼ 20%) increase of source luminosity, the parameters of the Comptonised spectral component would change drastically from the 1983 to the 1985 observation (a factor of > 10 and ∼ 0.1 variation in electron temperature and Thomson depth, respectively), when it would be virtually indistinguishable from a power law; (b) neutron star LMXRBs of luminosity comparable to that of 4U1957+11 do not show evidence for an additional black body component, with upper

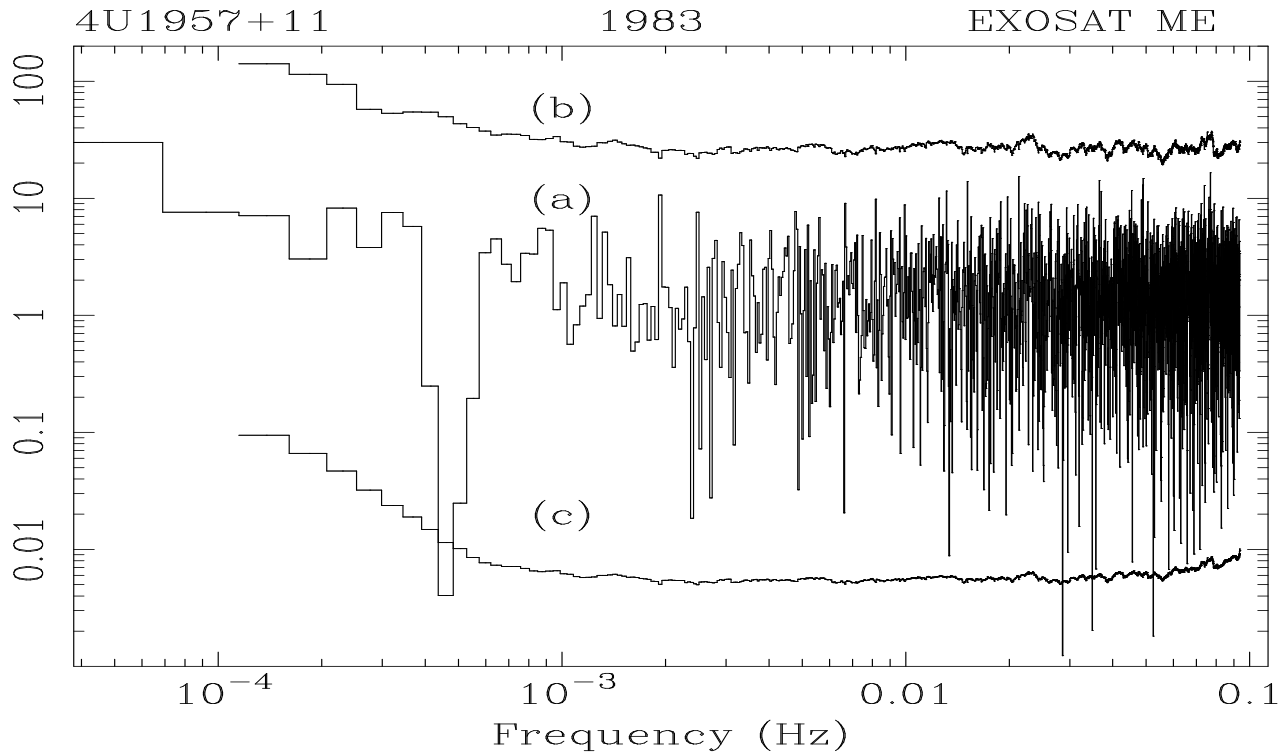

**Fig. 6.** Same as Figure 5, but for the 1–10 keV ME lightcurve of the 1983 August 27 observation.

limits usually in 10–20% range.

Black body emission disk models can be fit to the data in place of the cutoff power law model (or the Comptonisation model), but they require an unrealistic inclination of $i > 87.7°$ that cannot be reconciled with the absence of X-ray eclipses. This is also unlike other high state BHCs and a number of high luminosity LMXRBs (White et al. 1988).

A Fe K-shell emission feature is detected at an energy of $7.06 \pm 0.09$ keV in the 1985 data. The centroid energy of such a line is consistent only with $K\alpha$ transitions from H-like ions ($E_0 = 6.96$ keV) and is among the highest revealed with the *EXOSAT* GSPC (Gottwald & White 1991). $K\alpha$ transitions from any of the lower ionisation stages of iron require a substantial blueshift. Similar to the modelling of the Fe K-lines from other X-ray binaries and AGNs, such a blueshift could result from (relativistic) Doppler effects due to bulk plasma motions in the vicinity of the collapsed object. In particular, the observed feature might correspond to the blue horn of the characteristic profile produced in the innermost region of a relativistic accretion disk (Fabian et al. 1989). The line equivalent width ($\sim 100$ eV) is in the range measured from a number of LMXRBs (White et al. 1986). We note that the different line centroid energy derived by Singh et al.(1994) likely results from their drastically different modelling of the continuum.

The *EXOSAT* light curves of 4U1957+11 show only a moderate variability on timescales of a few hours, similar to BHCs in their high state (e.g. LMC X–3; Treves et al. 1988, Ebisawa 1991). A search for periodicities revealed no coherent modulation.

*Acknowledgements.* We are grateful to Dr. S. Ilovaisky for helpful comments on an earlier version of this paper. The EXOSAT spectra and light curves were obtained through the High Energy Astrophysics Database Service at the Brera Astronomical Observatory and the on-line interactive analysis system at the Space Science Department of ESA, ESTEC. This work was partially supported by ASI grants.

*Note added in proof.* In the recently distributed preprint "EXOSAT GSPC iron line catalog" by Gottwald et al. (1994, A&A, in press), no iron line detection is reported for the 1985 GSPC spectra of 4U1957+11, despite the fact that the continuum model is the same as that used in our paper. This is surpising since we have verified that the detection condition of Gottwald et al. (a $\chi^2$ improvement of $\geq 13$) is met also for the energy range used in their paper. We can speculate that perhaps their automatic fitting procedure failed to reach the minimum $\chi^2$ in the case of 4U1957+11.